\documentclass[5p,twocolumn,times,number]{elsarticle}
\usepackage{amsmath}
\usepackage{amssymb}
\usepackage{graphics}
\usepackage{graphicx}
\usepackage{epsfig}
\usepackage{dcolumn}
\usepackage{bm}
\usepackage{lineno}

\begin{document}
\begin{frontmatter}

\title{Measurement of spark probability of GEM detector for CBM muon chamber (MUCH)}
\author[label1,label2]{S.~Biswas\corref{cor}}
\ead{saikat.ino@gmail.com, s.biswas@niser.ac.in, saikat.biswas@cern.ch}
\author[label1]{A.~Abuhoza}
\author[label1]{U.~Frankenfeld}
\author[label1]{C.~Garabatos}
\author[label1]{J.~Hehner}
\author[label1]{V.~Kleipa}
\author[label1]{T.~Morhardt}
\author[label1]{C.~J.~Schmidt}
\author[label3]{H.~R.~Schmidt}
\author[label3]{J.~Wiechula}

\cortext[cor]{Corresponding author}

\address[label1]{GSI Helmholtzzentrum f\"ur Schwerionenforschung GmbH, Planckstrasse 1, D-64291 Darmstadt, Germany}
\address[label2]{School of Physical Sciences, National Institute of Science Education and Research, Jatni - 752050, India}
\address[label3]{Eberhard-Karls-Universit\"at, T\"ubingen, Germany}

\begin{abstract}
The stability of triple GEM detector setups in an environment of high energetic showers is studied. To this end the spark probability in a shower environment is compared to the spark probability in a pion beam. 

\end{abstract}
\end{frontmatter}

\section{Introduction}\label{intro}
The Compressed Baryonic Matter (CBM) experiment at the future Facility for Antiproton and Ion 
Research (FAIR) in Darmstadt, Germany, will use proton and heavy ion beams to study matter at 
extreme conditions and to explore the QCD phase diagram in the region of high baryon densities 
\cite{CBM,FAIR,CBM2008,APFAIR}. This will only be possible with the application of advanced 
instrumentation, including highly segmented and fast gaseous detectors such as Gas Electron 
multipliers (GEM) which will be employed in the Muon detector (MUCH) of the CBM experiment \cite{FS97}. 
The current design of the muon system consists of 6 iron absorber layers of thickness 20, 20, 20, 
30, 35, 100 cm respectively, interleaved with 6 detector stations, which will allow for tracking 
through the absorber stack.
The muon detector in the CBM experiment will be constructed in such 
a way that there will be micro-pattern gaseous detectors with high rate capability at least 
in the first 4 stations and other detectors like pad chambers, thick GEM or straw tubes in the 
later stations \cite{APFAIR}. A schematical view of the CBM detector with its MUCH detection 
system is shown in Figure~\ref{CBM}.

\begin{figure}
\begin{center}
\includegraphics[scale=0.25]{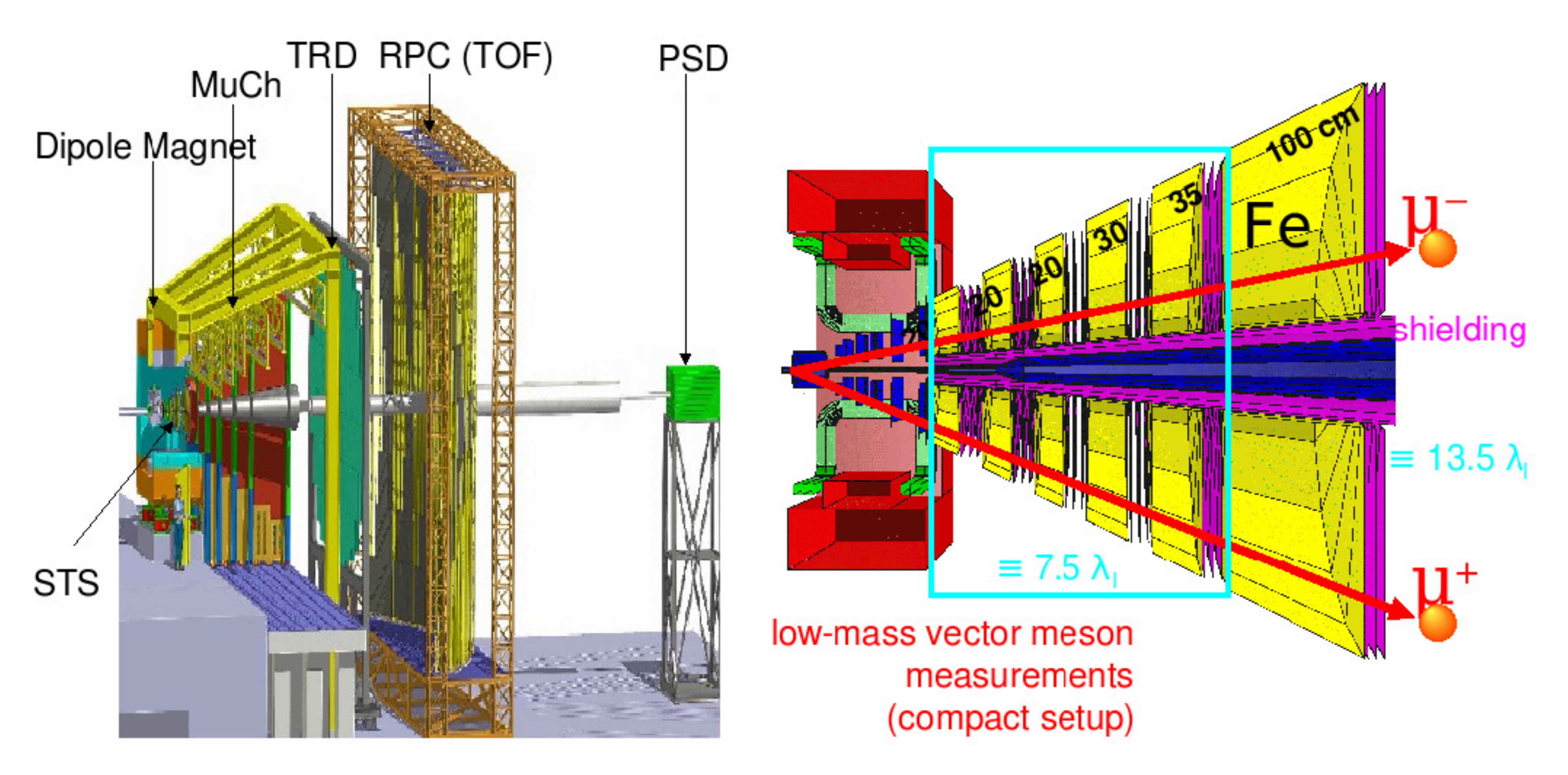}
\caption{\label{CBM} Schematic view of the CBM experiment: Muon set up (left). Implementation of 
the muon detection system in GEANT (right).}\label{CBM}
\end{center}
\end{figure}

Systematic studies on GEM detectors has been performed and published previously 
\cite{CBM11,SB11,SB12}. The main goal of this particular study is to measure the discharge 
probability of GEM detectors \cite{MC, MCA03} in a heavy shower environment that is to be expected for
the first stations of the CBM-MUCH. In a dedicated beam time double mask triple 
GEM detectors were tested at high rate at CERN SPS/H4 with the focus on the spark probability 
due to slow ionizing particles. To this end a pion beam of $\sim$150~GeV 
was used. Different methods to identify the occurrence of a spark are 
elaborated and the overall results on the spark probability measurement are presented.
\section{Description of the GEM module}\label{construct}

In this measurement double-mask GEMs in tripple stack configuration of 10~cm $\times$ 10~cm area 
were used. 
The drift gap, the two transfer gaps and the induction gap were kept at 2~mm. The read-out 
plane consist of 256 pads of 6$\times$6 mm$^2$ size. All the readout pads were routed to 
two connectors of 128 pins each. Even though the readout was segmented for the chamber, 
in this study the signals obtained from all pads 
were fed into a single channel charge sensitive preamplifier and analyzed with PXI LabVIEW 
based data acquisition system \cite{NI}. 
During the entire beam time the detector was operated with a premixed counting gas mixture 
of Argon and CO$_2$ in 70/30 ratio. 

\begin{figure}[htb!]
\begin{center}
\includegraphics[scale=0.37]{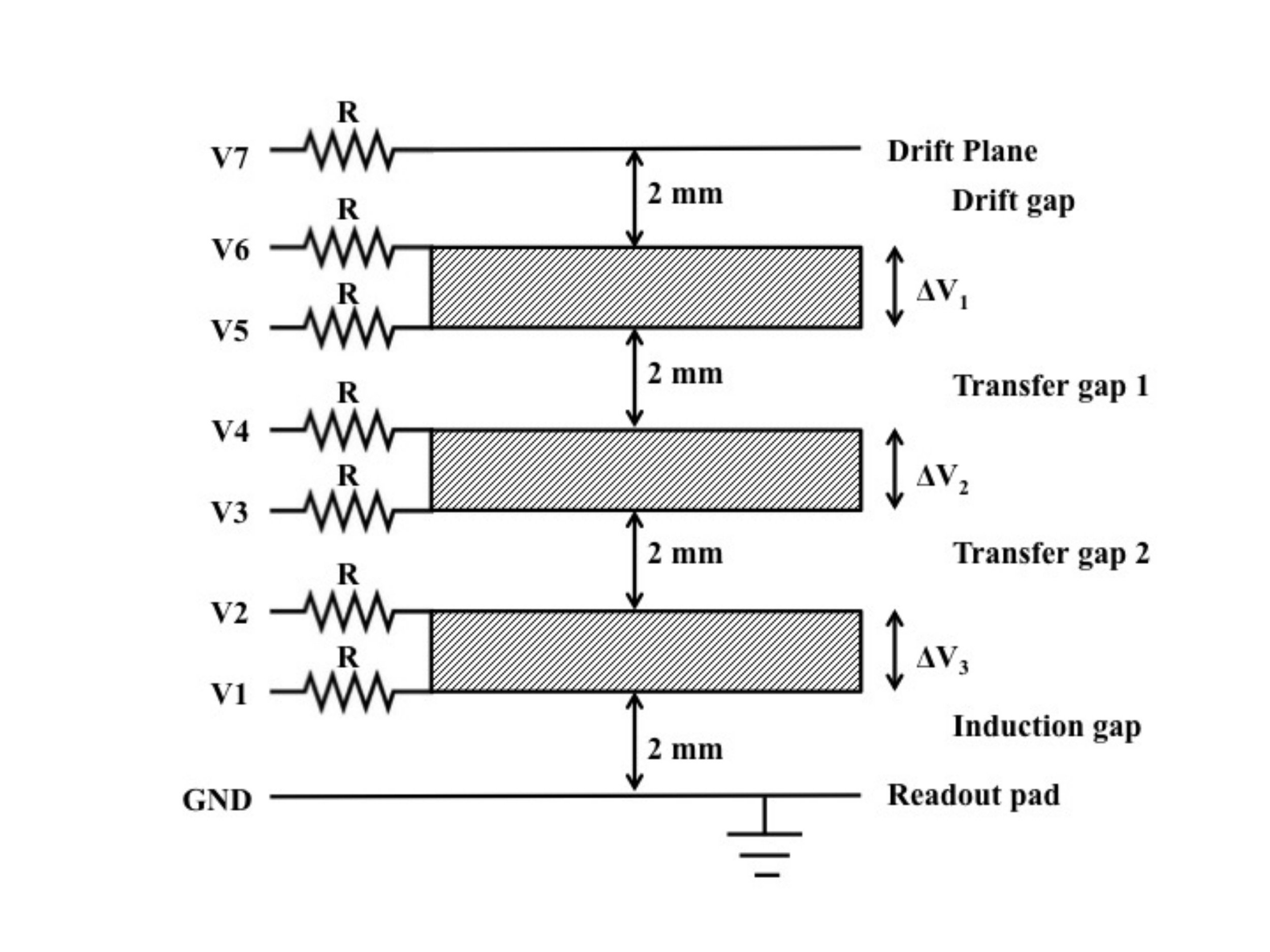}
\caption{\label{voltage}Voltage distribution in GEM.}\label{voltage}
\end{center}
\end{figure}

The operating potentials to the GEMs was applied by a seven-channel HVG210 
power supply made by LNF-INFN \cite{GC}. 
This module allows for controlling the supply voltages of a triple GEM detector. 
The module communicates with peripherals via CAN bus. 
The HVG210 power supply comprises seven almost identical channels, each 
of them being able to produce a specified voltage level with a current reading and 
current limiting option. 
The currents of all channels were recorded and used to determine the occurance of a spark. 
The configuration of potentials in the GEM detector is shown in Figure~\ref{voltage}. 
11 M$\Omega$ $[R]$ protection resistors were employed in all the seven channels. 
The details of the voltages and electric fields in the drift, induction and 
two transfer gaps are summarized in Table~\ref{tab:table1} \cite{SB12}.

\begin{center}
\begin{table}[htb!]
\begin{center}
\caption{Typical applied voltages and fields on the various gaps of a triple GEM chamber, 
operated with argon and CO$_{2}$ in a 70/30 mixing ratio.} \label{tab:table1}
\begin{tabular}{|c|c|c|c|} \hline
Gap Name & Gap width (mm) &  Voltage (V)  & Field (kV/cm) \\ \hline
Drift & 2 &  500  & 2.5 \\ \hline
Transfer 1 & 2 &  600  & 3.0 \\ \hline
Transfer 2 & 2 &  600  & 3.0 \\ \hline
Induction & 2 &  400  & 2.0 \\ \hline
\end{tabular}\\
\end{center}
\end{table}
\end{center}

\section{Arrangement for beam test}\label{setup}

Figure~\ref{shower} schematically shows the experimental set-up for measurement and identification 
of a spark due to a shower in the GEM detector using pion beam. Showers are provoked through a 
10~cm thick iron absorber placed into the primary beam. Four scintillators, two placed upstream 
the absorber into the primary beam and another two placed downstream the absorber, laterally shifted 
away from the beam direction, 
are used for obtaining a trigger from a shower created in the absorber. The four-fold 
coincidence between
signals from finger scintillators ScF1 and ScF2, and the two scintillators ScM and ScL 
is used to identify the presence of a shower. Upstream ScF1 
and ScF2 two additional scintillator detectors were allocated, whose coincidence signal monitors the 
rate of the primary beam from SPS. These detectors are referred to as the beam counter. 
The coincidence count of 
ScF1 and ScF2 proved to always match with the beam counter from SPS. 
The GEM-detector under investigation was placed off axis downstream the absorber 
outside the primary beam direction such that beam particles not having produced a shower 
would not go through the GEM-detector. Therefore, only those primary and secondary particles involved 
in a shower could reach the detector. Such particles are to a large extend highly ionizing particles.

\begin{figure}[htb!]
\begin{center}
\includegraphics[scale=0.3]{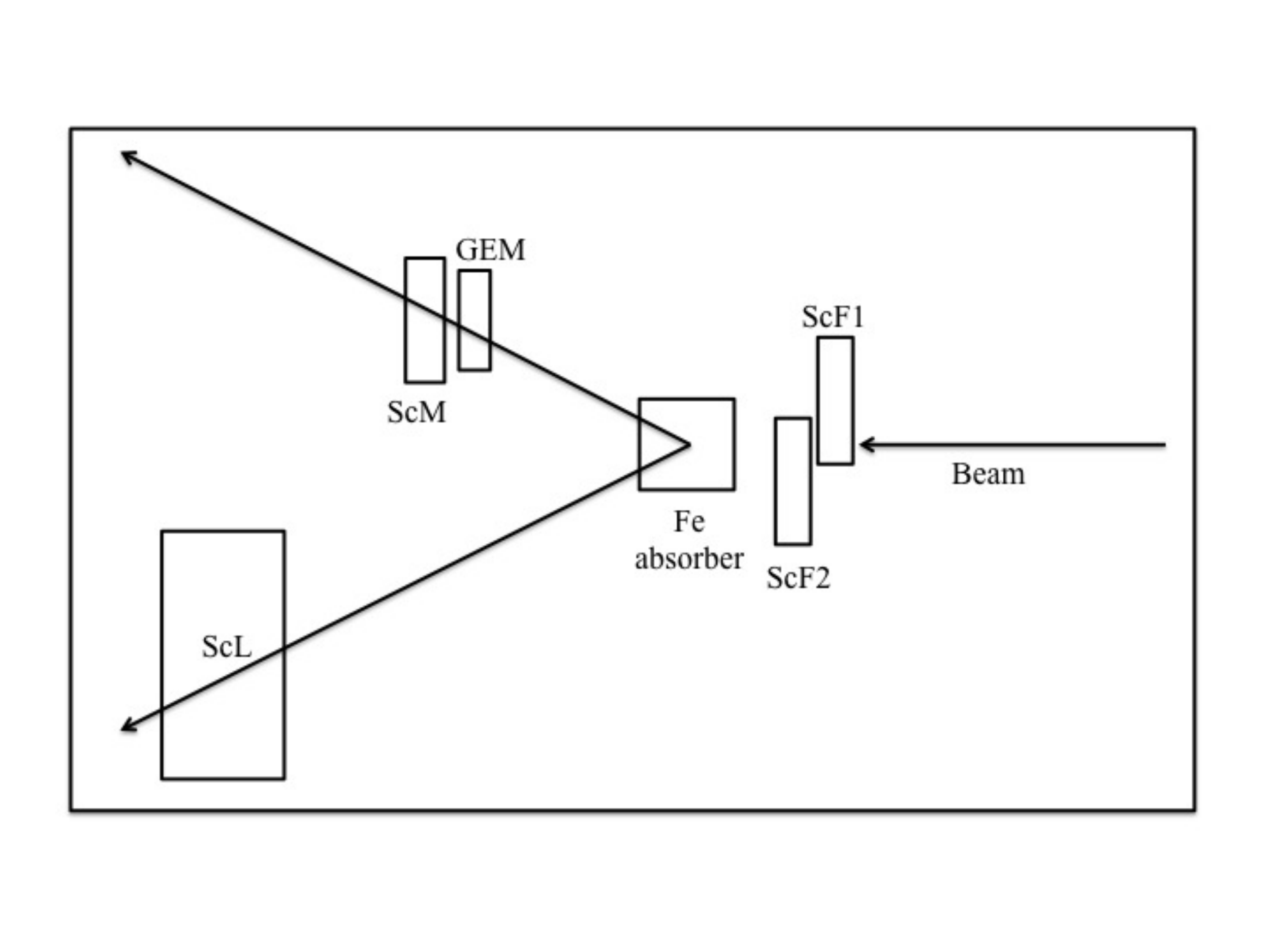}
\caption{\label{shower}Experimental set-up for the measurement of shower induced sparks. 
The dimension of the iron absorber was 10~cm~$\times$~10~cm~$\times$~20~cm. The GEM-detector was 
placed 16~cm laterally away from the primary beam direction at a downstream distance of 30~cm (entrance window) 
from the front surface of the iron absorber. 
}\label{shower}
\end{center}
\end{figure}
\begin{figure}[htb!]
\begin{center}
\includegraphics[scale=0.3]{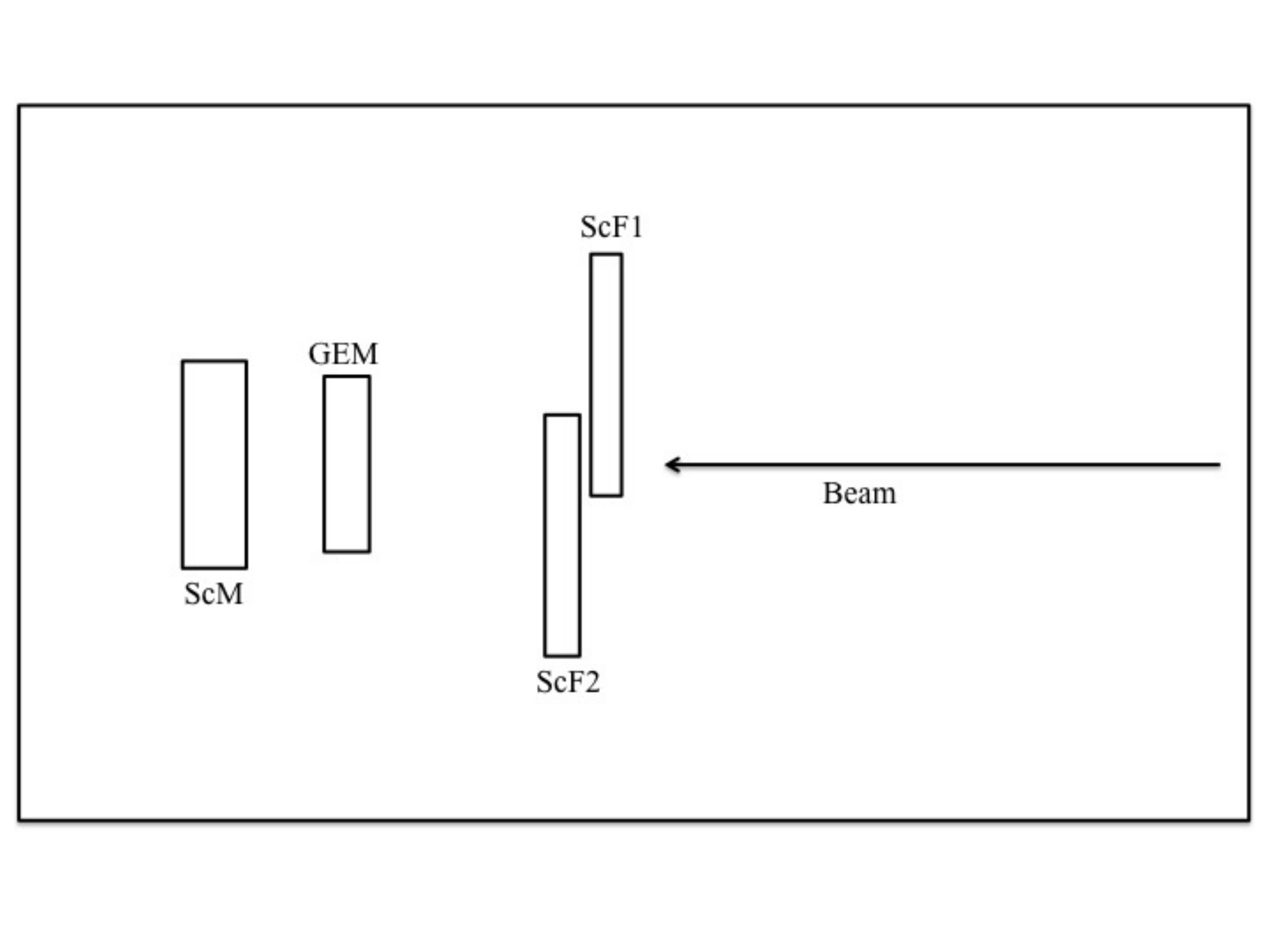}
\caption{\label{pion}Experimental set-up for the reference measurement of spark probability 
due to a pure pion beam.}\label{pion}
\end{center}
\end{figure}
\begin{figure}[htb!]
\begin{center}
\includegraphics[scale=0.35]{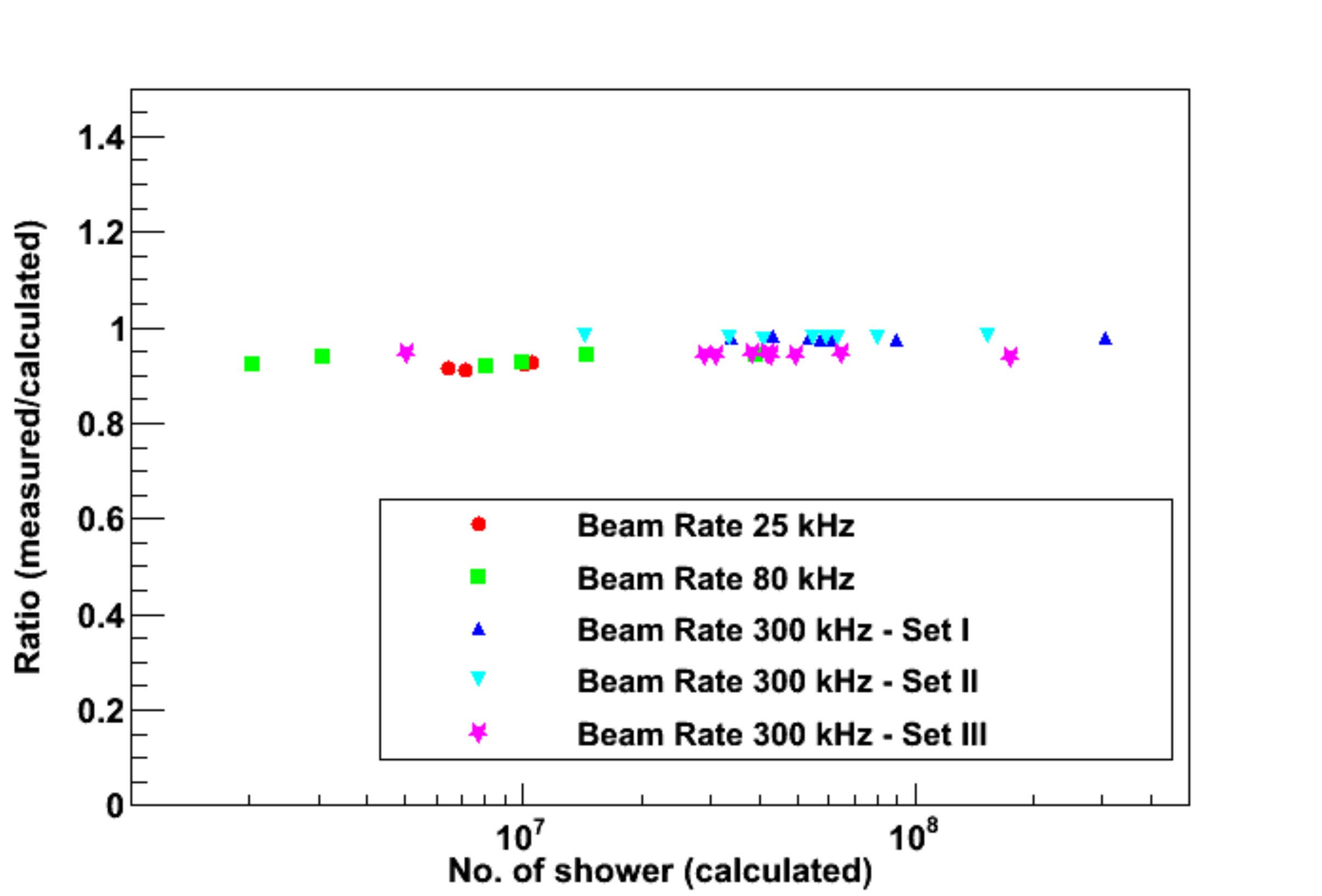}
\caption{\label{particle_comparison} Ratio of experiemtally determined shower intensity to FLUKA 
simulated expectations.}\label{particle_comparison}
\end{center}
\end{figure}

The experimental setup to measure the spark probability for the pion beam is shown in Figure~\ref{pion}. In this setup the three-fold coincidence between signals from ScF1, ScF2, placed in front of the GEM and ScM, placed after the GEM is taken to define a beam particle. In this case all the scintillators and the GEM are aligned with the beam line.

In this set-up the voltage and current from all seven channel of the HVG210, counts from the beam counter, scintillators, GEM detector and the pulse height of the GEM detector has been measured. In the result section we shall discuss only those measurements which are relevant to spark probability measurement.

The number of particles produced during shower has been calculated from the exact geometry of the experimental set-up by using FLUKA simulation. From the FLUKA simulation it was found that in the GEM plane the total number of electrons and positrons, neutrons, pions, protons, kaons and muons were respectively 0.8, 0.2, 0.08, 0.04, 0.006 and 0.001 per primary pion. The numbers are compared with the experimental value. The comparison for the different data set is shown in Figure~\ref{particle_comparison}. FLUKA simulations have shown that showers hitting the GEM area contain more then 40\% slow particles (mainly protons and pions) $\beta\gamma$ $<$2.

\section{Result}\label{res}

\subsection{Measurement of current}\label{current}

In this study the current in all the individual channels from the drift plane, and top and bottom surfaces of the three GEM foils has been recorded by the HVG210 power supply. The variation of current during and in between spills in SPS are shown in Figure~\ref{rate80kHz} and Figure~\ref{rate300kHz} respectively for pion beam and shower along with the GEM count rate. In the first case the average pion beam rate is 80~kHz and in the second case i.e., during shower the pion beam rate is 300~kHz. In both cases the voltage setting for the GEM were $\Delta V_1$=400~V, $\Delta V_2$=395~V and $\Delta V_3$=390~V. For 10 cm thick iron absorber around 35\% of the pion beam particles produced shower. The SPS has a spill of 10 seconds with an interval of about 30 seconds.

\begin{figure}[htb!]
\begin{center}
\includegraphics[scale=0.45]{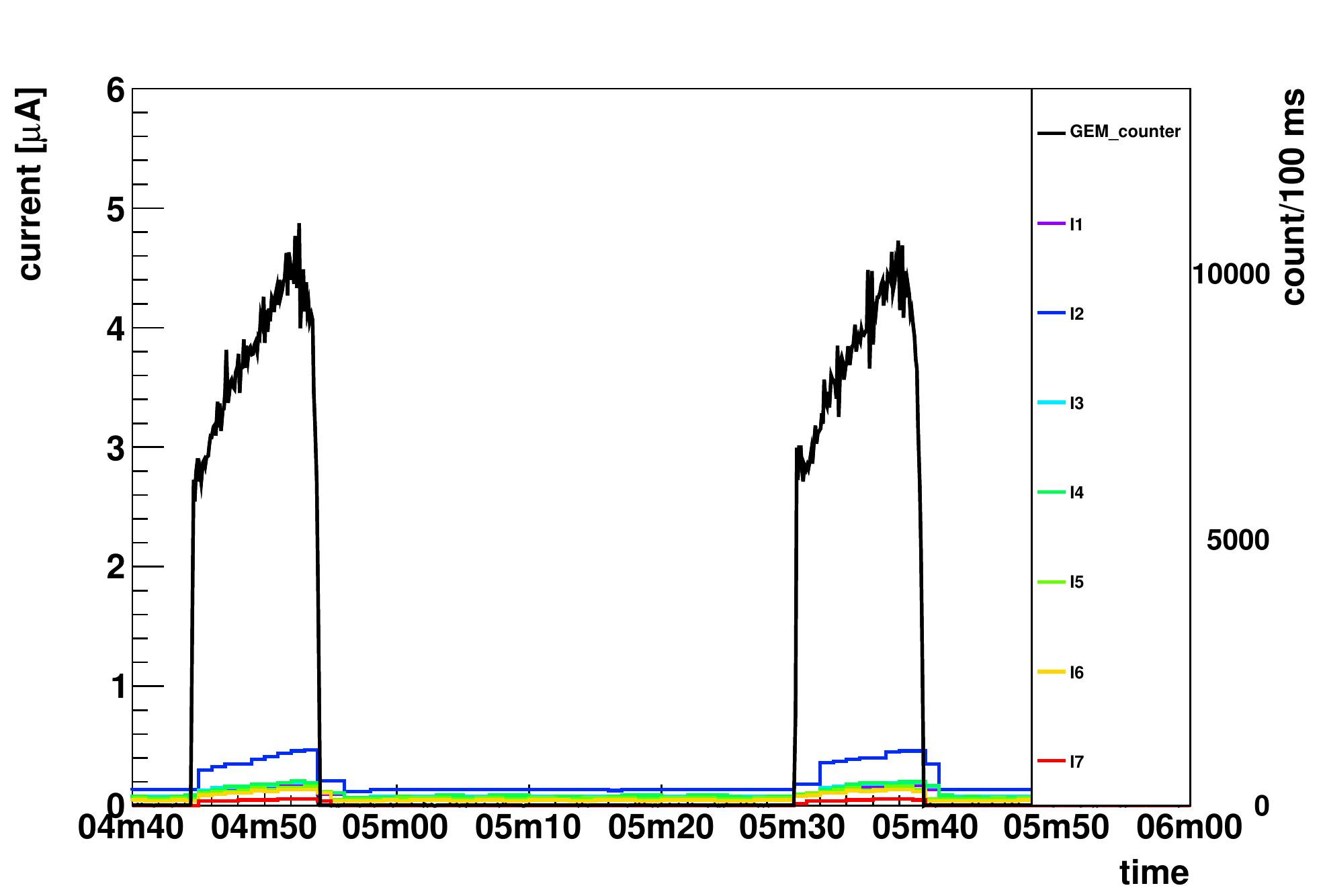}
\caption{\label{rate80kHz}Current and GEM counting rate: Pion beam 80 kHz.}\label{rate80kHz}
\end{center}
\end{figure}
\begin{figure}[htb!]
\begin{center}
\includegraphics[scale=0.45]{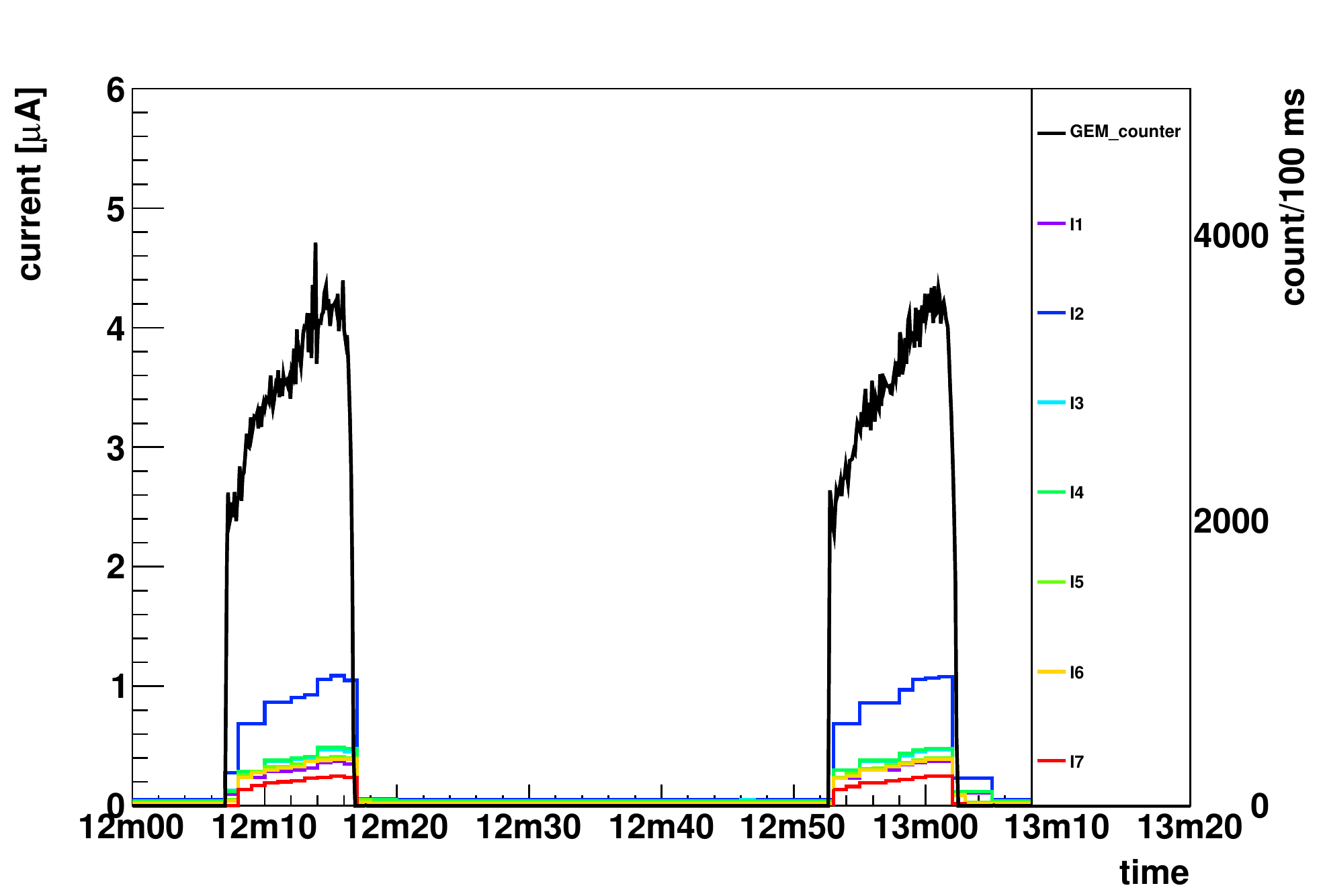}
\caption{\label{rate300kHz}Current and GEM counting rate during Shower: Beam rate300 kHz.}\label{rate300kHz}
\end{center}
\end{figure}

\subsection{Measurement of spark probability}\label{current}
The spark probability is defined as the ratio of the total number 
of sparks produced in the GEM detector to the total number of particles incident 
on the detector \cite{SB01,GB02}.

In this paper two different methods to identify the presence of a spark in the GEM-detector were used.
The first one identifies a spark through the absence of a signal, the detector being saturated. 
The second method identifies the presence of a spark through an enhancement of operating currents.
Both these methods were experimentally executed in two different ways: 
Absence of signal was identified through a drop in the count rate on one side and the absence 
of signals in the pulse height analyzing ADC. The second method through observation of 
operating currents can identify a spark either through a jump in a particular current or through an 
internal "spark counter" of the power supply. 
These methods are discussed in detail in the following paragraphs.

Among these different methods the first and most important one proved to be the identification
of a spark from a sudden drop of counting rate in the GEM detector during a spill. 
During a spark, a sudden drop in the electric field inside a GEM reduces the overall gain and thus the 
counting rate in the GEM-detector. A situation where the ratio of count rates between GEM-detector 
and beam counter drops below 20\% of its average value is defined as a spark.  Different 
threshold values between 10\% and 50\% were tested, but no significant change of the results 
was observed. This definition appears as a rather robust way to identify the presence 
of a spark in the GEM-detector during a spill and was thus employed for this experimental 
investigation.   

In Figure~\ref{onespark} the black line exemplary shows the count rate registered by the GEM-detector 
during a spill. The apparent sudden drop in count rate is used to define the occurance 
of a one spark. Even the presence of two sparks during a spill was sometimes observed. Figure~\ref{twospark} shows such an example.

\begin{figure}[htb!]
\begin{center}
\includegraphics[scale=0.4]{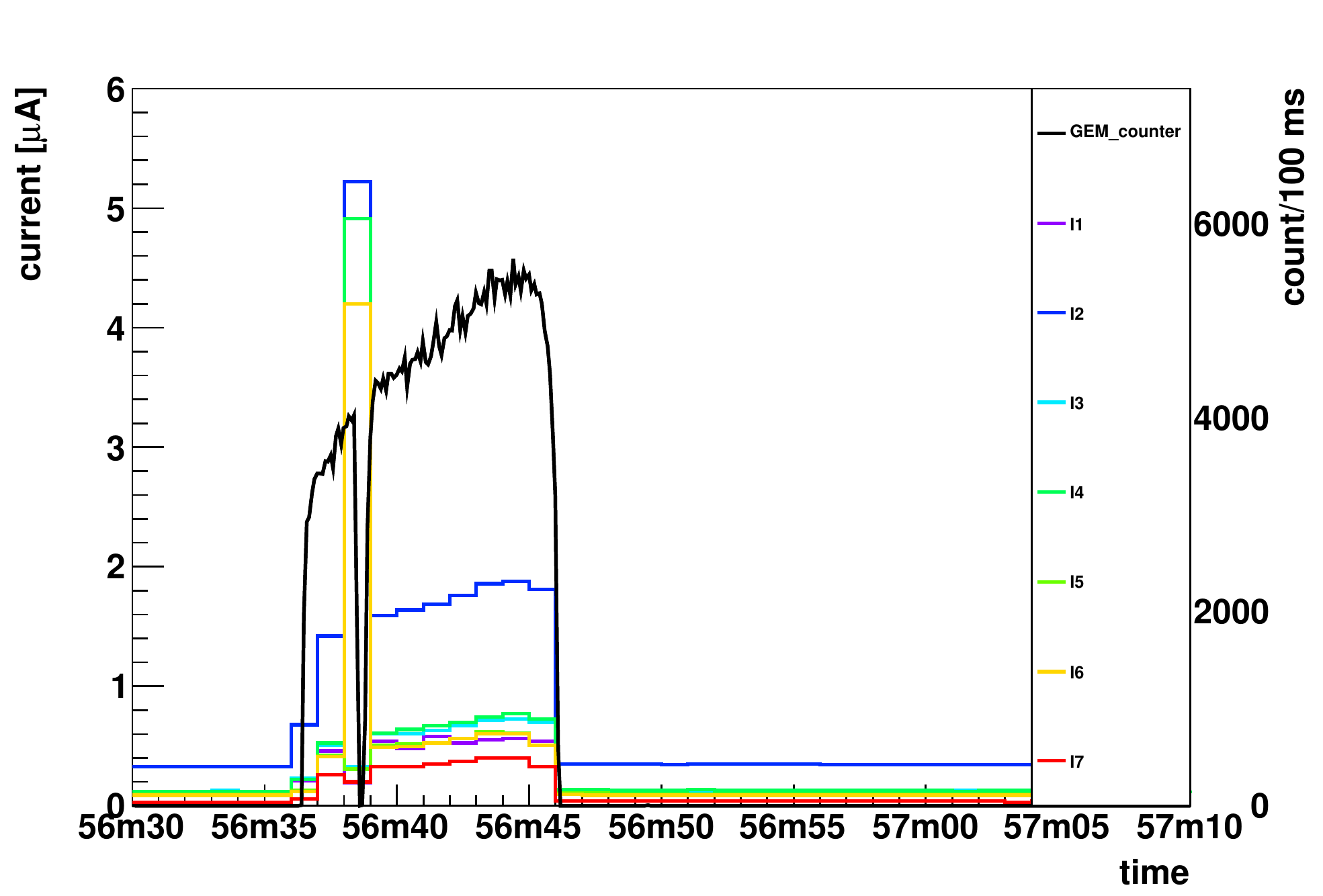}
\caption{\label{onespark}Identificaton of a spark from a drop in count rate in the GEM during 
a spill. In parallel, the currents on all GEM-electrodes were registered and are displayed.}\label{onespark}
\end{center}
\end{figure}
\begin{figure}[htb!]
\begin{center}
\includegraphics[scale=0.4]{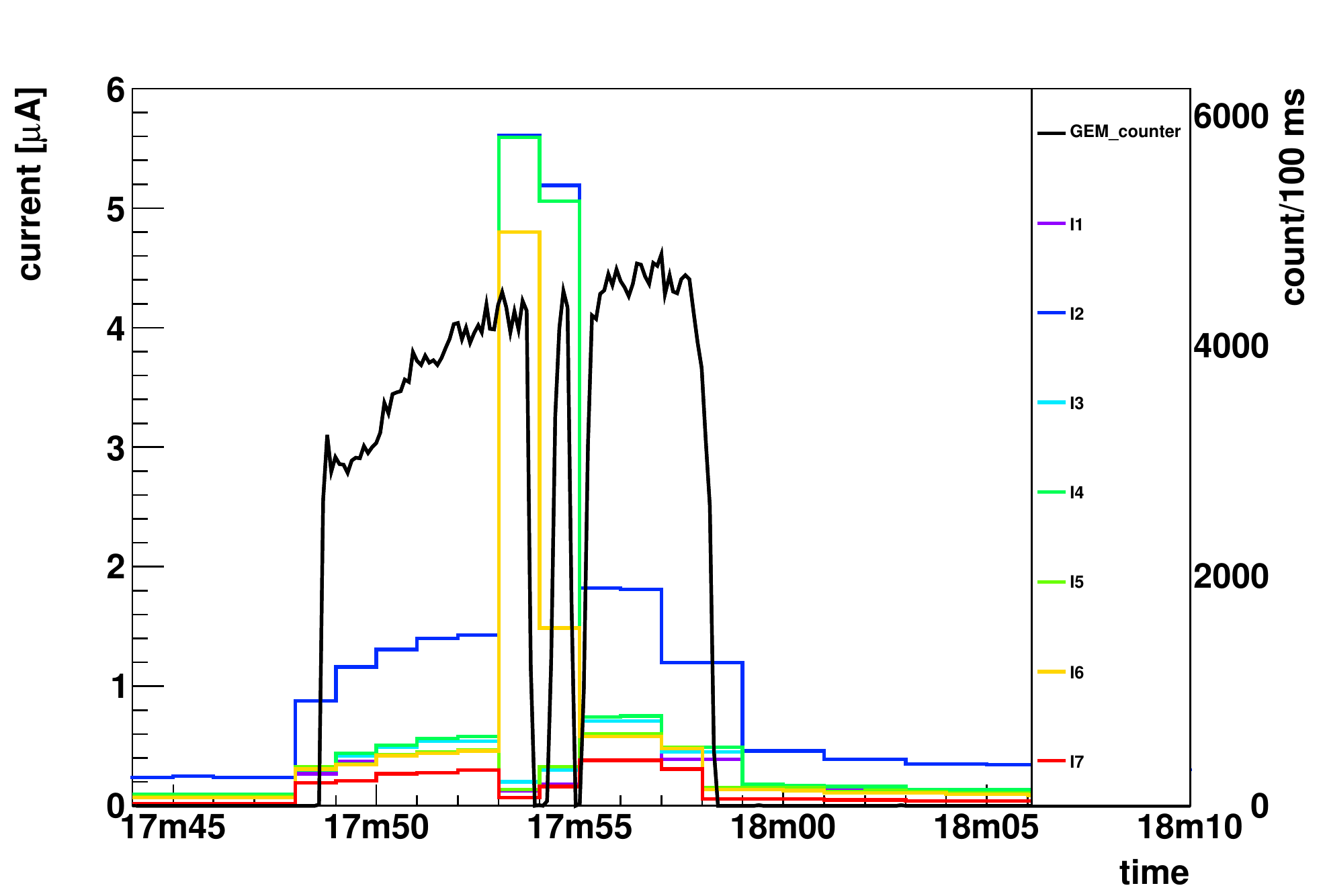}
\caption{\label{twospark}Example of a spill with the occurance of two sparks.}\label{twospark}
\end{center}
\end{figure}

The second method to determine a spark uses the sampling ADC data. When there is a spark 
inside the GEM the electric field drops resulting no signal from the GEM. Thus, in case of a 
spark in the GEM-stack,
one should not exect a signal on the ADC channel observing the GEM signal, 
which on itself is triggered through the coincidence signal of the scintillators. 
The use of this method depends upon the noise level to be as small as possible and for 
this reason proved to be much less robust for the identification of a spark.  

Measuring the current from the individual channels and observing an increase 
in the current value particularly from the top of the three GEMs is another way 
to determine the occurance of a spark during a spill. The effectiveness can be seen in figures~\ref{onespark} 
and \ref{twospark}. Whenever there is a drop in the count rate in the GEM-detector, 
some or all of the current values in the power supply show a steep jump from their normal values. 
Thus, the occurrence of a spark can alternatively be determined from such a sudden current increase
on any one of the GEM-electrodes. The current threshold in our experiments was set 
to 4~$\mu$A to define a spark.

The latter method to detect a spark can be implemented through a built-in feature of the 
GEM power supply HVG210. The module comprises a special provisions called trip checker 
to count a spark. A current threshold I$_{th}$ can be set together with a time window T$_{wnd}$ and
the number of times (N$_{ovfl}$) the current readings has gone beyond 
the specified threshold value, so that the module internally registers the occurance of a spark. 
With a high current threshold value, a count of the N$_{ovfl}$ yields an estimate the number 
of sparks that occured. An increase in N$_{ovfl}$ indicates in increase in the number of sparks. 
This method has one particular advantage: It yields the number of sparks in a particular 
supply channel and thus points to the particular GEM foil where the spark occured.\\

The determined spark probability as a function of the global GEM voltage 
($\Delta V_1$ + $\Delta V_2$ + $\Delta V_3$), the sum of the potential differences of every GEM in the stack,
 is shown in 
figure~\ref{sparkprob} (cf. \cite{peskov14,SB2012}). The measurement was taken with 
different gas flow rates, 3 lt/hr and 5 lt/hr. 
In earlier measurements it had been observed that the gain of the detector depended upon the 
gas flow rate \cite{SB11}. In this operational global voltage range the gain of the detector
 was measured to vary between 12000 and 30000. In this figure the spark probability is 
determined through the first method only i.e. from the drop in the GEM count rate during a spill. 
In this study the overall spark probability was found to be $\sim 10^{-7}$. The probable reason 
for this high absolute value is the operation of the GEM-detector at very high 
gain \cite{SB99, SB02}. Two different measurements i.e. a jump in the current and a drop in the 
counting rate yield almost identical results. During off-spill the spark probability was practically 
zero.\\
The spark probability during operation of the GEM detector in showers does not increase if compared
to the operation in a pure pion beam. Thus, in such a relative measurement, relating the 
suszeptibility to sparks in a shower environment to the spark probability in a pion beam, the overall 
spark probability that may depend upon the experimental particularities, is not that relevant.\\
The showers apparently really do put
some load on the GEM-detector as it is found in FLUKA simulation. This can be confirmed through the
higher electrode currents in the stack during a shower. Yet, the spark probability is not compromized 
through these heavy, highly ionizing showers.

\begin{figure}[htb!]
\begin{center}
\includegraphics[scale=0.45]{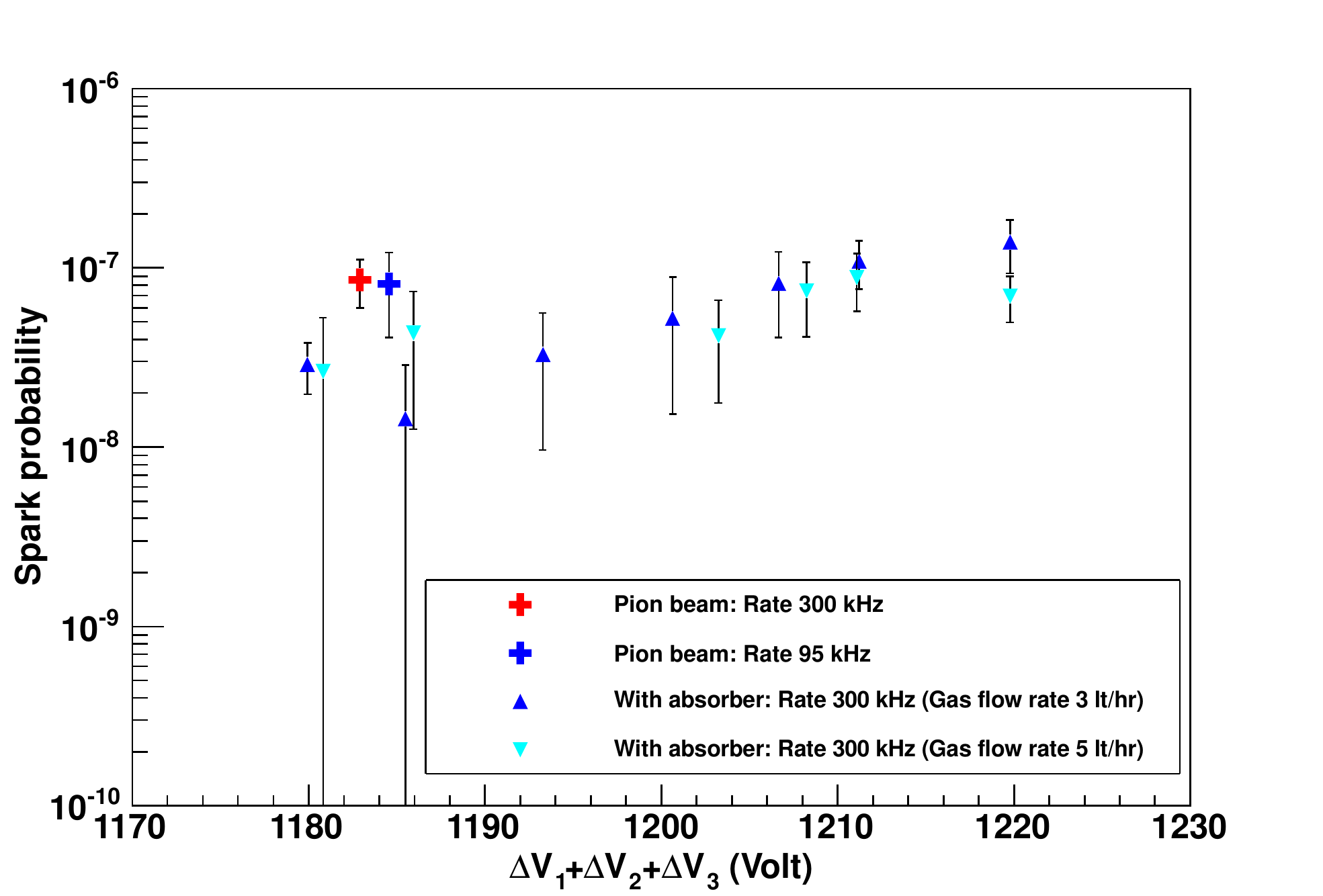}
\caption{\label{sparkprob}Spark Probability as a function of global GEM voltage.}\label{sparkprob}
\end{center}
\end{figure}

\section{Conclusions and outlooks}
In this article four different methods for the identification of spark in a GEM detector 
were elaborated. The spark probability was measured for a pure pion beam as well as a pion beam 
a with 10~cm thick iron brick as absorber. The variation of the spark probability as a 
function of the global GEM voltage ($\Delta V_1$ + $\Delta V_2$ + $\Delta V_3$) has been 
presented for a shower and for the pion beam. 
The spark probability increases exponentially with the global voltage. In this study the 
spark probability was found to be $\sim 10^{-7}$ at the applied voltage settings. A secondary shower does not appear to trigger additional sparks in 
the GEM detector when compared to a pure pion beam. It is concluded that the presence of slow, heavily ionizing particles contained in showers behind the iron absorbers does not increase the spark probability relative to fast pions.
Current measurements on the GEM electrodes show an incrase of the current in allmost all the electrodes of 
the stack. The maximum absoulte increase in current is observed in the third GEM-foil, where 
the maximum number of electrons is reached. The foils, in which the spark occurs, 
can be identified through the separately measured currents.

As an outlook, the measurement will be repeated with 3~mm drift gap and without a current limiting protection 
resistor at the bottom of the GEMs in future. The effects will be studied and will be 
communicated at a later stage.

\section{Acknowledgements}
We are thankful to Dr. Ingo Fr\"{o}hlich of University of Frankfurt, 
Prof. Dr. Peter Fischer of Institut f\"{u}r Technische Informatik der Universit\"{a}t Heidelberg, 
Prof. Dr. Peter Senger, CBM Spokesperson and Dr. Subhasis Chattopadhyay, Deputy spokesperson, CBM  for their support in course of this work. We are grateful to Dr. Anna Senger of GSI for the FLUKA simulation results. We would like to thank Jorrit C. L. Widder for his effort during the SPS test beam. We are also grateful to Dr. Leszek Ropelewski and Dr. Serge Duarte Pinto of RD51 for their valuable suggestions. S. Biswas acknowledges the support of DST-SERB Ramanujan Fellowship (D.O. No. SR/S2/RJN-02/2012).

\end{document}